\documentclass[pdf,preprint]{iucr}              % DO NOT DELETE THIS LINE
\usepackage{graphicx}
\usepackage{amsmath}
\paperprodcode{a000000}      % Replace with production code if known
\paperref{xx9999}            % Replace xx9999 with reference code if known
\papertype{CP}               % Indicate type of article
\paperlang{english}          % Can be english, french, german or russian
\journalcode{J}              % Indicate the journal to which submitted
\journalyr{2012}
\journaliss{1}
\journalvol{68}
\journalfirstpage{000}
\journallastpage{000}
\journalreceived{0 XXXXXXX 0000}
\journalaccepted{0 XXXXXXX 0000}
\journalonline{0 XXXXXXX 0000}

\begin{document}                  % DO NOT DELETE THIS LINE

\title{PASCal: A principal-axis strain calculator for thermal expansion and compressibility determination}
\shorttitle{PASCal: A principal-axis strain calculator}

\author{Matthew J.}{Cliffe}
\cauthor{Andrew L.}{Goodwin}{andrew.goodwin@chem.ox.ac.uk}

\aff{Department of Chemistry, University of Oxford, Inorganic Chemistry Laboratory, South Parks Road, Oxford, OX1 3QR \country{U.K.}}

\shortauthor{Cliffe and Goodwin}

\maketitle                        % DO NOT DELETE THIS LINE

\begin{synopsis}
The web-based program {\sc pasc}al (http://pascal.chem.ox.ac.uk) is introduced as a tool for determining principal coefficients of thermal expansion and compressibilities using variable-pressure and variable-temperature lattice parameter data. The additional scientific insight provided by this analysis is illustrated with reference to three case studies: the metal--organic framework Cu-SIP-3, the molecular crystal $\beta$-HMX and the mineral malayaite.
\end{synopsis}

\begin{abstract}
We describe a web-based tool ({\sc pasc}al; {\bf p}rincipal {\bf a}xis {\bf s}train {\bf cal}culator, http://pascal.chem.ox.ac.uk) designed to simplify the determination of principal coefficients of thermal expansion and compressibilities from variable-temperature and variable-pressure lattice parameter data. In a series of three case studies, we use {\sc pasc}al to re-analyse previously-published lattice parameter data and show that additional scientific insight is obtainable in each case. First, the two-dimensional metal--organic framework Cu-SIP-3 is found to exhibit the strongest area-negative thermal expansion (NTE) effect yet observed; second, the widely-used explosive HMX exhibits much stronger mechanical anisotropy than had previously been anticipated, including uniaxial NTE driven by thermal changes in molecular conformation;  and, third, the high-pressure form of the mineral malayaite is shown to exhibit a strong negative linear compressibility (NLC) effect that arises from correlated tilting of SnO$_6$ and SiO$_4$ coordination polyhedra.
\end{abstract}

\section{Introduction}

The mechanisms by which crystalline materials respond to changes in temperature and pressure are as important for the valuable insight they provide into the fundamental chemistry of the solid state as they are for a range of practical applications. Topical examples include negative thermal expansion (NTE) in metal--organic frameworks (MOFs) \cite{Wu_2008,Lock_2010}, negative linear compressibility (NLC) in molecular co-crystals \cite{Fortes_2011} and extreme gating responses to guest sorption in microporous materials \cite{Serre_2002,Mulfort_2010,Walton_2011}. From an experimental viewpoint, crystallographic techniques offer a direct and straightforward measure of any structural variation: the evolution of unit cell parameters describes the bulk mechanical response of the material, with the corresponding changes in atomic positions indicating the microscopic mechanism responsible for this response. So, for any given system, determination of the temperature- and/or pressure-dependence of its unit cell parameters is a simple but effective means of diagnosing anomalous (and useful) mechanical behaviour.

\begin{figure}\label{fig1}
\begin{center}
\caption{An illustrative example of the effect of unit cell choice on lattice parameter expansivity determination: (a) A hypothetical two-dimensional molecular framework, with both a rhombic and an orthogonal unit cell indicated; (b) a plot of the relative change in lattice parameter determined for each of these two cells for the concerted lattice deformation indicated by arrows in (a). The rhombic cell shows no variation in unit cell length whereas the changes in orthogonal cell lengths reflect the underlying bulk material response.}
\includegraphics{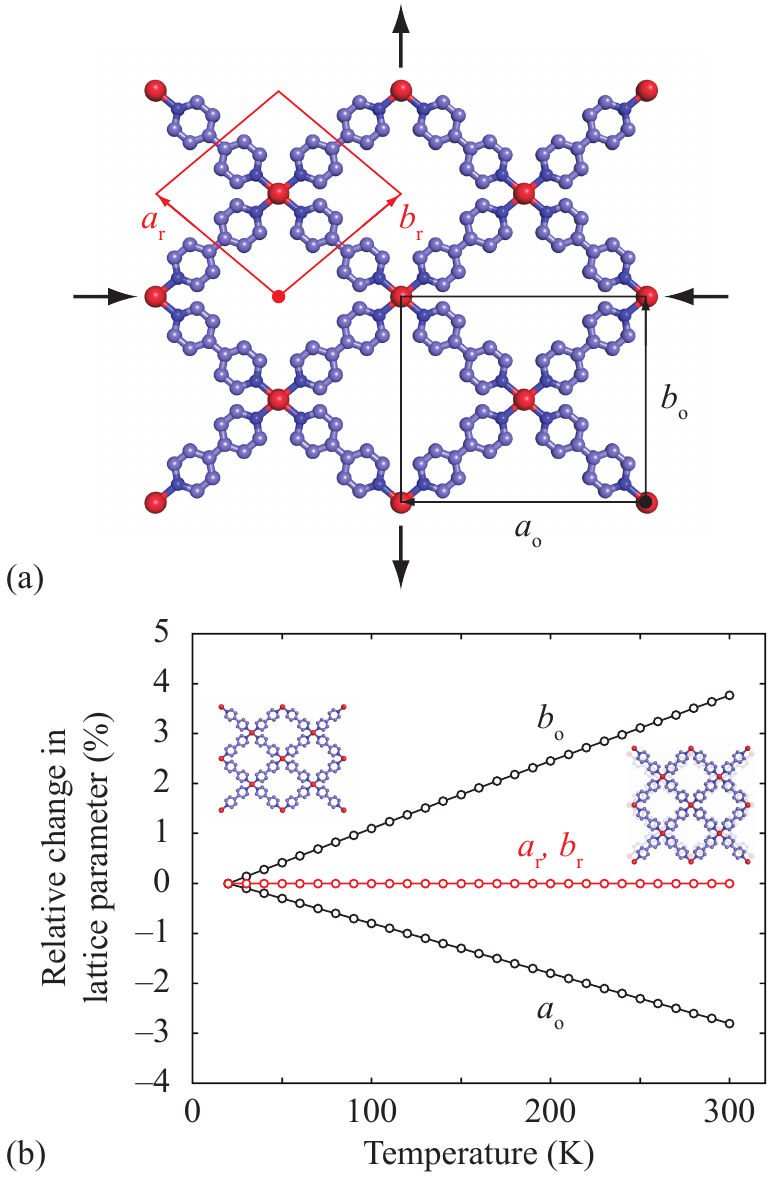}
\end{center}
\end{figure}

Because the unit cell is a construct of convention rather than uniqueness, the question as to whether variations in unit cell geometry actually reflect the fundamental mechanical response\footnote{In this manuscript we use the term `mechanical response' to refer to those mechanical properties most readily measured crystallographically: hydrostatic compressibilities, the bulk modulus, the pressure derivative of the bulk modulus, and the thermal expansivities. In other contexts, the same term is sometimes equated with the full elastic tensor and includes \emph{e.g.}\ the elastic compliances, Poisson's ratio and shear moduli \cite{Nye_1957}.}  of a material is an important one. Put differently: can we be certain that an alternate choice of unit cell---however unconventional---would yield the same lattice expansivities and/or compressibilities? The answer, of course, is that the values determined nearly always depend on the particular unit cell chosen. A canonical example is given by a simple two-dimensional molecular framework [Fig.~\ref{fig1}(a)] (a lower dimensional analogue, perhaps, of MIL-53 \cite{Serre_2002} or Ag$_3$[Co(CN)$_6$] \cite{Goodwin_2008}). The dominant deformation mechanism of this framework might involve a `wine-rack'-like expansion in one direction that couples to a contraction in a perpendicular direction \cite{Baughman_1998}---a mechanism that is reflected in the lattice parameter variation of the orthogonal (rectangular) unit cell shown in Fig.~\ref{fig1}.  In that of the related rhombic cell [Fig.~\ref{fig1}(b)], there is no apparent change in the values of the lattice parameters $a$ and $b$; instead the important variation occurs in the angle $\beta$, which is less readily interpreted. By analogy, in real materials of low crystal symmetry the key science can be hidden within the variation of the entire ensemble of lattice parameters $a,b,c,\alpha,\beta,\gamma$, making analysis much less straightforward than would otherwise be hoped (or indeed is often assumed).

The tensor algebra that resolves this issue is well established but is perhaps used more widely within some communities than others. The central concept is that for any system there exists a unique set of orthogonal axes (the principal axes) along which the material responds in a purely linear fashion---either expanding or contracting---with the response along other directions involving some non-zero shear component. It is the expansion or contraction along the principal axes that describes the fundamental mechanical response of the system. Predictably, crystal symmetry plays a crucial role in determining the orientation of these axes. For systems of orthorhombic symmetry or higher the principal and (conventional) crystallographic axes coincide \cite{Nye_1957}.\footnote{For uniaxial systems (\emph{i.e.}, with 3-, 4- or 6-fold rotational symmetry), two of the three principal axes are related by this rotational symmetry. These axes define a unique plane in which the mechanical response is invariant. Cubic systems are isotropic.} Hence the variation in lattice parameters $\ell$ with temperature and pressure can be related directly to the bulk expansivities $\alpha_\ell$ and compressibilities $K_\ell$:\footnote{There is more than one convention widely in use for the symbols for compressibility and its inverse the bulk modulus. In key solid-state physics texts (\emph{e.g.}\ \citeasnoun{Ashcroft_1976} and \citeasnoun{Kittel_2005}) $K$ is used to denote compressibility and $B$ the bulk modulus. Confusingly, the primary reference for mechanical properties of crystals \cite{Nye_1957} uses $K$ to denote the bulk modulus and $\beta$ the compressibilities. Here, we follow the physics convention in order to remove any ambiguity regarding the lattice parameter $\beta$.}
\begin{eqnarray}
\alpha_\ell&=&+\frac{1}{\ell}\left(\frac{\partial\ell}{\partial T}\right)_p,\label{alpha}\\
K_\ell&=&-\frac{1}{\ell}\left(\frac{\partial\ell}{\partial p}\right)_T.\label{k}
\end{eqnarray}
For monoclinic and triclinic crystal classes, however, the principal axes are not necessarily related to the unit cell axes\footnote{For monoclinic classes in the conventional setting, one principal axis lies parallel to $\mathbf b$.} and must instead be determined as the eigenvectors of the full strain tensor. In particular, because the principal expansivities and compressibilities now depend in a complex fashion on the variation in unit cell lengths and angles, the values determined using Eqs.~(\ref{alpha}) and (\ref{k}) no longer carry any physical significance. Returning to the example of Fig.~\ref{fig1}, the discrepancy in lattice thermal expansivities arises because the principal axes coincide with the axes of the rectangular unit cell rather than the rhombic cell.

Variable-temperature (or -pressure) unit cell data contain all the information necessary to identify the principal axes and hence the principal coefficients of thermal expansion (or compressibilities)---but for low-symmetry systems the process of calculating these is a problem best suited to automation. A number of software packages are capable of carrying out this calculation: notably {\sc strain} \cite{Ohashi_1982} and its successor {\sc winstrain},\footnote{The {\sc winstrain} program is unpublished but available for download from www.rossangel.com.} {\sc deform} \cite{Filhol_1987} and {\sc {e}}l{\sc {am}} \cite{Marmier_2010}. Yet for much of the materials chemistry community, the many features offered by these packages can in some cases require more detailed experimental data than are available (\emph{e.g.}\ elastic constant measurements), and/or provide a level of analysis beyond what is necessary to identify expansivity and compressibility anomalies.

We have developed a new web-based tool, {\sc pasc}al (= {\bf p}rincipal {\bf a}xis {\bf s}train {\bf cal}culatorl; available at http://{\sc pasc}al.chem.ox.ac.uk) that accepts variable-temperature or variable-pressure lattice parameter data as input and simply returns as output the principal-axis expansivities/compressibilities and the orientation of the principal axes relative to the crystallographic axes. The pressure-dependence of the unit cell volume is also fitted to equations of state, from which the bulk modulus and its pressure derivative are determined. This output enables the user to identify quickly any anomalous mechanical properties (\emph{e.g.}\ NTE, NLC, pressure-induced softening) and also to understand their geometric relationship to the structure itself. At all times, our particular focus has been on providing a simple tool useable by experts in elastic property calculations and non-experts alike.

Our paper is arranged as follows:  in Section \ref{computation}, we describe the particular implementation of strain tensor calculations used by the program and provide a summary of the most important aspects of the program input and output. What follows subsequently is a series of three case studies---focusing on one MOF, one molecular crystal, and one mineral---where we have used {\sc pasc}al to re-analyse previously-published lattice parameter data. In all cases we are able to develop the relevant science as a result of considering the principal compressibilities or expansivities. In this way, we hope to demonstrate that routine application of {\sc pasc}al to parametric crystallographic studies is not only feasible, but that the program is capable of adding substantive scientific value to the field.

\section{Computational Detail}\label{computation}

\subsection*{Theory}
The calculations at the heart of {\sc pasc}al involve determination of the orthogonal strains. Our starting point is the transformation from crystallographic axes $\mathbf A_i$ to orthogonal axes $\mathbf E_i$ (with notation as in Giacovazzo {\it et al.} (2011)). The orientation of the $\mathbf E_i$ is arbitrary: we make use of the Institute of Radio Engineers convention where $\mathbf E_3$ is parallel to the $\mathbf c$ crystallographic axis, $\mathbf E_1$ is parallel to $\mathbf a^\ast$, and $\mathbf E_2=\mathbf E_3\times\mathbf E_1$. The corresponding change-of-basis transformation is described by the square matrix $\mathbf M$:
\begin{equation}
\mathbf E=\mathbf M\times\mathbf A.
\end{equation}
The strain $\boldsymbol\epsilon$ corresponding to a pair of initial and final conditions (\emph{i.e.}, temperature or pressure values) is given as the symmetric part of the product of $\mathbf M$ at the first condition with its inverse at the second condition: defining
\begin{equation}
\mathbf e\equiv\mathbf M_{\textrm{final}}^{-1}\times\mathbf M_{\textrm{initial}}-\mathbf I,
\end{equation}
we have
\begin{equation}
\boldsymbol\epsilon=\frac{1}{2}(\mathbf e+\mathbf e^{\textrm T}).
\end{equation}
 The eigenvalues $\epsilon_i$ and eigenvectors $\mathbf x_i$ of the matrix $\boldsymbol\epsilon$ are then the principal strains and the principal axes. The derivatives of these principal strains with respect to temperature and/or pressure give the principal coefficients of thermal expansion and/or compressibilities:
\begin{eqnarray}
\alpha_i&=&+\left(\frac{\partial\epsilon_i}{\partial T}\right)_p\\
K_i&=&-\left(\frac{\partial\epsilon_i}{\partial p}\right)_T\label{kcalc}
\end{eqnarray}
Formally, the definitions we have given here correspond to infinitesimal Lagrangian strains---this is the standard approach applied in texts such as \citeasnoun{Nye_1957}. We note however that infinitesimal strains are quantitatively exact only in the limit of small variations in lattice parameters. Consequently, slightly different definitions of strain can be more meaningful in other situations (\emph{e.g.}\ finite strain for lattice discontinuities at a phase transition). There is scope within {\sc pasc}al to calculate these alternative strains; interested readers are referred to \citeasnoun{Giacovazzo_2011}, \citeasnoun{Schlenker_1978} and \citeasnoun{Zotov_1990} for further details regarding the choice of strain and the associated calculations. In practice the use of different strain definitions can give quantitatively significant differences in principal axis expansivity/compressibility values, but is unlikely to affect the diagnosis of qualitative features, such as the existence or otherwise of NTE or NLC. Implementation of all strain calculations within {\sc pasc}al is as a Fortran90 routine that makes use of the {\sc lapack} and {\sc minpack} libraries \cite{Anderson_1999,More_1984}.

\subsection*{Program input}
\begin{figure}\label{fig2}
\begin{center}
\caption{Sample {\sc pasc}al input for (a) variable-temperature and (b) variable-pressure lattice parameter data measured for Ag$_3$[Co(CN)$_6$]-II  \cite{Goodwin_2008}.}
\includegraphics{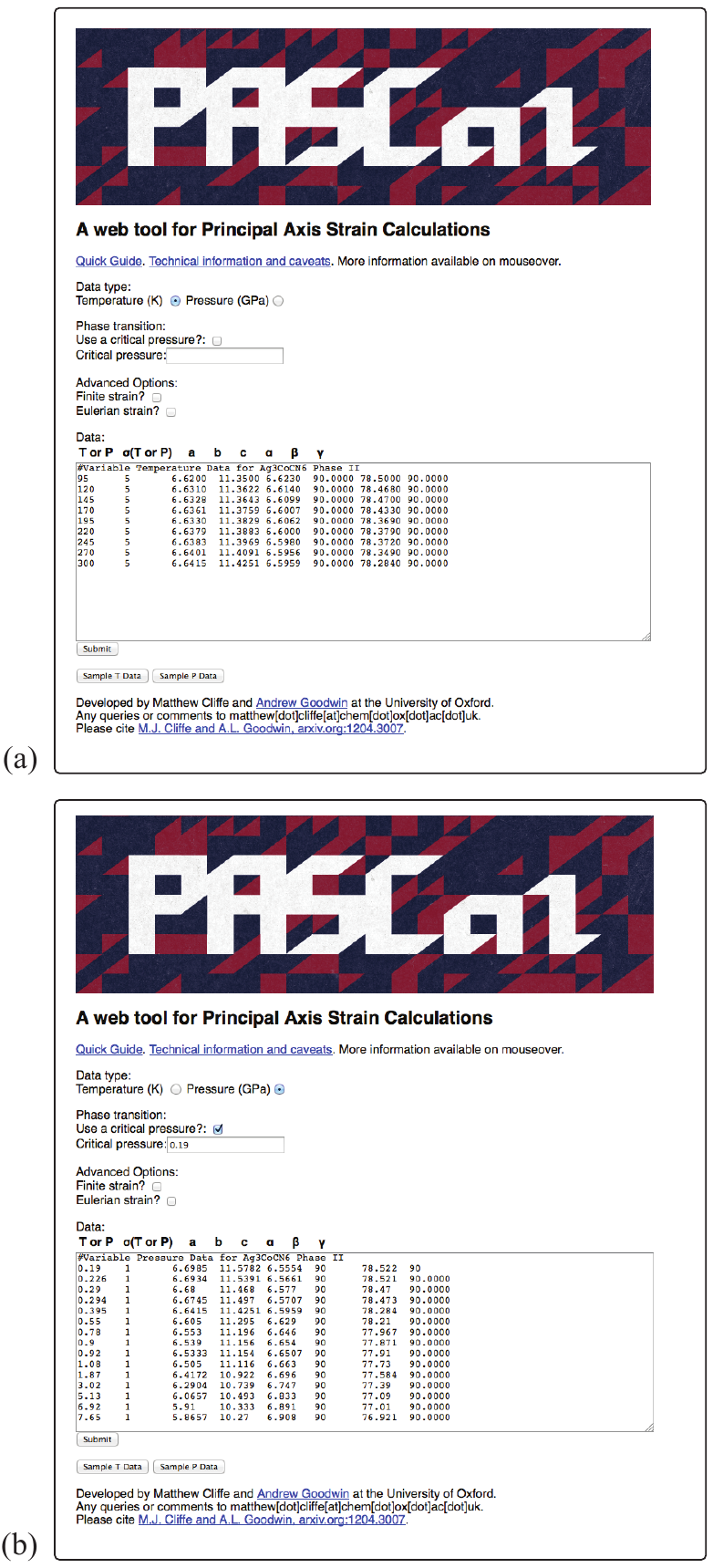}
\end{center}
\end{figure}

We have tried to simplify the required input for {\sc pasc}al as much as possible. Parametric data are entered as a series of whitespace- or comma-separated values in sequential columns: temperature (K) or pressure (GPa) values; the uncertainty in these values; followed by the corresponding lattice parameters $a, b, c, \alpha, \beta, \gamma$. At this stage, estimated errors in cell parameters are neither required nor taken into account, as the errors in all subsequently calculated parameters are nearly always dominated by errors in temperature or pressure values and scatter in the data themselves.\footnote{The use of standard errors on lattice parameters to determine standard errors in derived parameters has been the subject of a number of detailed investigations. First, it is well established that standard uncertainties are underestimated by both single-crystal and powder (Rietveld) refinement, with reported errors smaller by up to an order of magnitude than the real values determined by multiple measurements \cite{Taylor_1986,Herbstein_2000}. Second, the avoidance of bias in derived orthogonal cells arising from angular uncertainty is a well-recognised problem for monoclinic and triclinic systems \cite{Haestier_2009}. Third, the covariances necessary for accurate error propagation are rarely quoted or included in deposited CIFs. Some of these issues can be overcome with careful treatment---\emph{e.g.} by using repeat measurements to obtain experimental distributions of cell parameters---hence providing physically-meaningful estimates of standard uncertainties. A thorough investigation of one such approach is detailed in \citeasnoun{Angel_2000}. But, in order to maintain a simple interface and in order not to give an impression of misleading precision, the error-propagation algorithms within {\sc pasc}al have been implemented in such a way that is likely to provide a small overestimation of uncertainties. For specific instances demanding greater control over error treatment, the user is referred to more specialist code---\emph{e.g.}\ {\sc strain} \cite{Ohashi_1982}.} In Fig.~\ref{fig2} we show an example set of input data for both expansivity and compressibility measurements of the high pressure phase Ag$_3$[Co(CN)$_6$]-II \cite{Goodwin_2008}.

\subsection*{Program output}
For variable-temperature data, {\sc pasc}al calculates the principal thermal expansivities across the entire temperature range for which data are provided. No attempt is made to fit the expansivity data at any level of complexity beyond a simple linear dependence: $\ell_i(T)=\ell_0(1+\alpha_iT)$. More complex fitting approaches, such as using a Debye model for thermal expansion, are prone to instabilities in the absence of very high quality data that extend to low temperatures $T\ll\Theta_{\textrm D}$ \cite{David_1993}. We are of the view that complex fits are most likely to be of interest to specialist users, with simple linear coefficients of thermal expansion of greatest value to the broadest user base. In any case, the principal strains are given in tabular format for each temperature point, with the intention that more complex fits might easily be performed using any third-party fitting software of choice.

\begin{figure}\label{fig3}
\begin{center}
\caption{A comparison of three methods for representing an expansivity tensor, using the variable-temperature Ag$_3$[Co(CN)$_6$]-II data of \citeasnoun{Goodwin_2008}. (a) The thermal expansivity indicatrix, with PTE shown in red and NTE in blue;  (b) the thermal expansivity quadric; (c) the strain ellipsoid (shown here for the maximum strain obtained in order to emphasise the difficulty associated with perceiving asphericity). The directions and relative magnitudes of large PTE and NTE are obvious in the expansivity indicatrix representation, but are obscured in the other, more traditional, representation surfaces.
}
\includegraphics{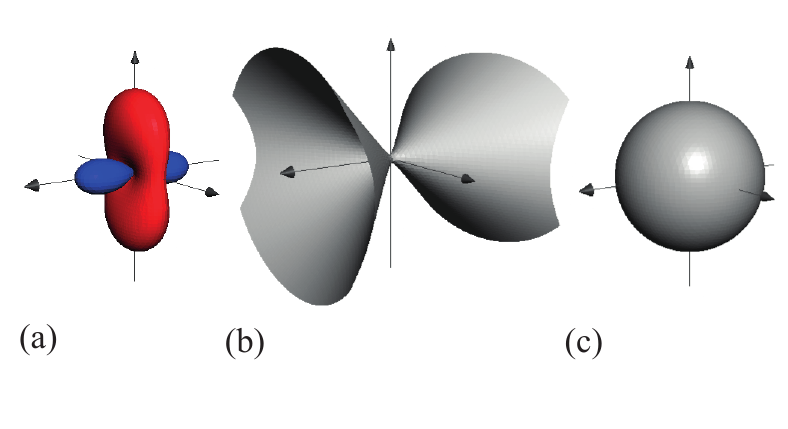}
\end{center}
\end{figure}

\begin{figure}\label{fig4}
\begin{center}
\caption{(a) The key {\sc pasc}al output obtained for the input data shown in Fig.~\ref{fig2}(a), showing the principal components of the expansivity tensor and the expansivity indicatrix representation as discussed in the text. (b) A representation of the crystal structure of Ag$_3$[Co(CN)$_6$]-II shown with axes in the same orientation as the expansivity indicatrix in (a). The layer of Ag$^+$ cations (shown as large spheres) corresponds to the large positive thermal expansion component of the expansivity indicatrix, which is shown in red; the perpendicular (stacking) axis of Ag$_3$[Co(CN)$_6$]-II then corresponds to the large NTE component of the expansivity indicatrix (shown in blue). The Co--C--N--Ag--N--C--Co linkages, shown in ball-and-stick representation, are oriented in the same set of directions for which the expansivity indicatrix has a near-zero value.}
\includegraphics{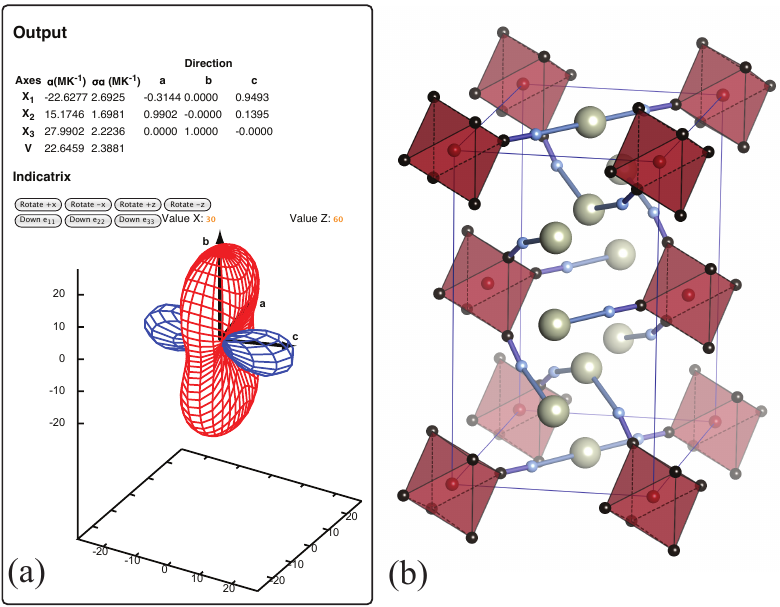}
\end{center}
\end{figure}

The relationship between principal axes and the crystallographic axes is also described in the {\sc pasc}al output, where we give the normalised components of the principal axes projected onto the crystallographic axes. The thermal expansivity tensor is represented visually in the form of the expansivity indicatrix. The expansivity indicatrix is a smoothly-varying surface centred at the unit cell origin, where the distance between the surface and the origin in a given direction $\mathbf r$ is equal to the magnitude of $\alpha$ in that same direction (taking negative values as appropriate) \cite{Belusov_2007,Durka_2011,Krashnenko_2000,Vershinin_2005}. Perhaps the more traditional method of visualising the expansivity or compressibility tensor would be \emph{via} either the representation quadric or the strain ellipsoid \cite{Nye_1957}. The particular advantage of the expansivity/compressibility indicatrix over these other representations---and the reason behind its inclusion in the {\sc pasc}al output---lies in the ease with which crystallographic directions corresponding to mechanical behaviour of particular interest can be identified (\emph{e.g.} $\alpha$ very large and positive, very large and negative, or zero). We illustrate this point explicitly in Fig.~\ref{fig3}, where we show together the expansivity indicatrix, the quadric and the strain ellipsoid for the one system Ag$_3$[Co(CN)$_6$]-II . For this material, comparison of the expansivity indicatrix with the crystal structure shows---in accord with the original study \cite{Goodwin_2008}---that NTE occurs along the layer-stacking axis and that the covalent Co--C--N--Ag--N--C--Co linkages are oriented along directions for which the expansivity essentially vanishes. The actual {\sc pasc}al output for this same system is shown in Fig.~\ref{fig4}.

\begin{figure}\label{fig5}
\begin{center}
\caption{Calculated fits to variable-pressure Ag$_3$[Co(CN)$_6$]-II lattice parameter data obtained using {\sc pasc}al: (a) relative changes in length for the three principal axes, (b) the corresponding principal compressibilities, and (c) the modified third-order Birch-Murnaghan fit to the unit cell volumes.}
\includegraphics{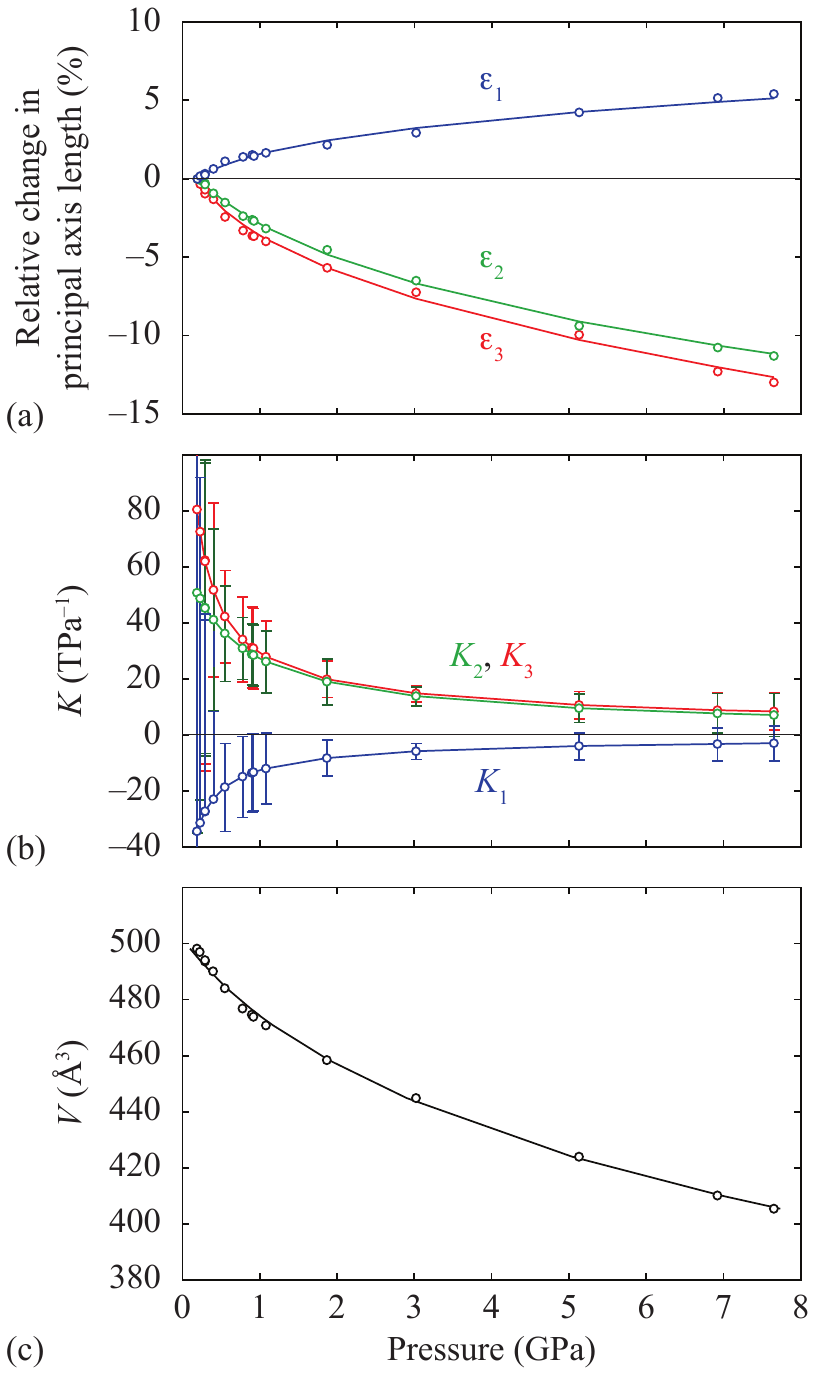}
\end{center}
\end{figure}

The output provided for variable-pressure data is more detailed. At the simplest level, median principal compressibilities, their uncertainties, and the corresponding principal axes are calculated according to Eq.~\ref{kcalc} using a linear fit to the strains. While thermal expansion data are usually approximately linear across typical measurement ranges, a linear pressure-dependence is rarely observed for variable-pressure data (and thermodynamically forbidden over large pressure ranges). In our experience of relatively highly-compressible materials such as molecular frameworks, axial compressibilities are often much better characterised using an empirical fit of the form $\ell(p)=\ell_0+\lambda(p-p_{\textrm c})^\nu$ \cite{Goodwin_2008,Cairns_2012}. Accordingly, {\sc pasc}al also outputs fitted values of $\lambda$ and $\nu$ for each principal axis and---because compressibility now varies with pressure---the $K_i(p)$ for each pressure point and their uncertainties are both plotted graphically and listed in tabular form [Fig.~\ref{fig5}].

There is valuable thermodynamic information in the pressure-dependence of the unit cell volume, in the form of the bulk modulus $B=(-V{\rm d}p/{\rm d}V)_T$ and its (dimensionless) pressure derivative $B^\prime=({\rm d}B/{\rm d}p)_T$. The parameter $B^\prime$ can be an indicator of unusual mechanical behaviour; for example, a large $B^\prime$ is often indicative of a rapid stiffening characteristic of layered materials \cite{Munn_1972} and a negative $B^\prime$ can indicate the presence of a dynamic instability \cite{Chapman_2007}.
Accordingly, unit cell volumes are fitted in {\sc pasc}al using Birch-Murnaghan equations of state \cite{Birch_1947} of various degrees of complexity, with the ultimate choice left to the user, perhaps guided by the quality of fits obtained or by construction of a so-called f--F plot \cite{Angel_2000}. Defining the parameter
\begin{equation}
\eta\equiv\left(\frac{V_0}{V}\right)^{\frac{1}{3}},\label{eta}
\end{equation}
where $V_0$ is the zero-pressure unit cell volume, the second- and third-order Birch-Murnaghan fits correspond to the equations of state
\begin{equation}
p(V)=\frac{3B}{2}\left(\eta^7-\eta^5\right)
\end{equation}
and
\begin{equation}
p(V)=\frac{3B_0}{2}\left(\eta^7-\eta^5\right)\left[1+\frac{3}{4}(B^\prime-4)(\eta^2-1)\right],\label{tobm}
\end{equation}
respectively. In Eq.~(\ref{tobm}), $B_0$ can be interpreted as the value of $B$ at zero pressure. For completeness, the option is provided to use finite Eulerian rather than infinitesimal Lagrangian strains, as the former is used in the derivation of the Birch-Murnaghan equation of state \cite{Birch_1947}. We note, however, that this formalism was derived for isotropic crystals and need not describe equally well the behaviour of anisotropic systems. Consequently Birch-Murnaghan fits for anisotropic systems should be interpreted with this caveat in mind.

For high-pressure phases that are unstable below some critical pressure $p_{\rm c}>0$, even the third-order Birch-Murnaghan equation of state does not capture the elastic behaviour fully, as it assumes that both $V$ and $B$ are continuous functions of $p$ for all $p>0$. In {\sc pasc}al, we have incorporated the option to fit a modified version, which now takes into account the existence of a non-zero critical pressure $p_{\rm c}$ \cite{Sata_2002}:
\begin{eqnarray}
p(V)&=&\eta^5\left[p_{\rm c}-\frac{1}{2}(3B-5p_{\rm c})(1-\eta^2)\vphantom{\frac{35p_{\rm c}}{9B_0}}\right.\nonumber\\
& &+\left.\frac{9}{8}B_0\left(B^\prime-4+\frac{35p_{\rm c}}{9B_0}\right)(1-\eta^2)^2\right].\label{bigeos}
\end{eqnarray}
In fitting Eq.~(\ref{bigeos}) the assumption is made that $p_{\rm c}$ is less than or equal to the lowest pressure value for which data are given. The user is given the option to include a fixed value of $p_{\rm c}$ in their analysis.

For the example of Ag$_3$[Co(CN)$_6$]-II discussed above and shown in Fig.~\ref{fig2}, it is meaningful to fix the value of $p_{\rm c}$ at the phase-I/II transition pressure of 0.19\,GPa. A good fit to the experimental $V(p)$ data is then obtained by {\sc pasc}al, as shown in Fig.~\ref{fig5}. The fundamental parameters we extract in this fit show some discrepancy with those reported in \citeasnoun{Goodwin_2008}: here we obtain $B_0=14.2(14)$\,GPa and $B^\prime=10.9(13)$ (the previously-reported values are 11.8(7)\,GPa and 13.5(12), respectively). The origin of this discrepancy appears to be the extremely large parametric covariances within the third-order Birch-Murnaghan equation of state resulting in different convergence behaviour with different minimisation routines (the values reported in \citeasnoun{Goodwin_2008} were determined using the solver routine within Microsoft Excel). A fit using the program {\sc eos-fit} \cite{Angel_2001} (which automatically constrains $p_{\rm c}=0$, and hence gives a slightly-different set of parameters yet again) returns covariance terms with magnitudes greater than 0.94, indicating that the system of equations is very poorly constrained. It is actually the case for many well-studied systems that there exists a variety of $B$, $B^\prime$ and $V_0$ values that result in very similar fits to experimental compressibility data, and we reiterate here the caution of \emph{e.g.}\ \citeasnoun{Jackson_1998} by emphasising the covariance amongst these parameters, especially for data with large scatter.

\section{Case Studies} 
\subsection{A metal--organic framework: Cu-SIP-3}
\begin{figure}\label{fig6}
\begin{center}
\caption{Negative thermal expansion in Cu-SIP-3$\cdot$3H$_2$O. (a) A single covalently-bonded layer in the material, viewed down the $[101]$ axis (the single positive thermal expansion axis). (b) The general mechanism responsible for area-NTE in layered materials: the lowest energy deformations correspond to `rippling' of the layers, which results in an increased interlayer separation but a smaller effective layer area.}
\includegraphics[width=7cm]{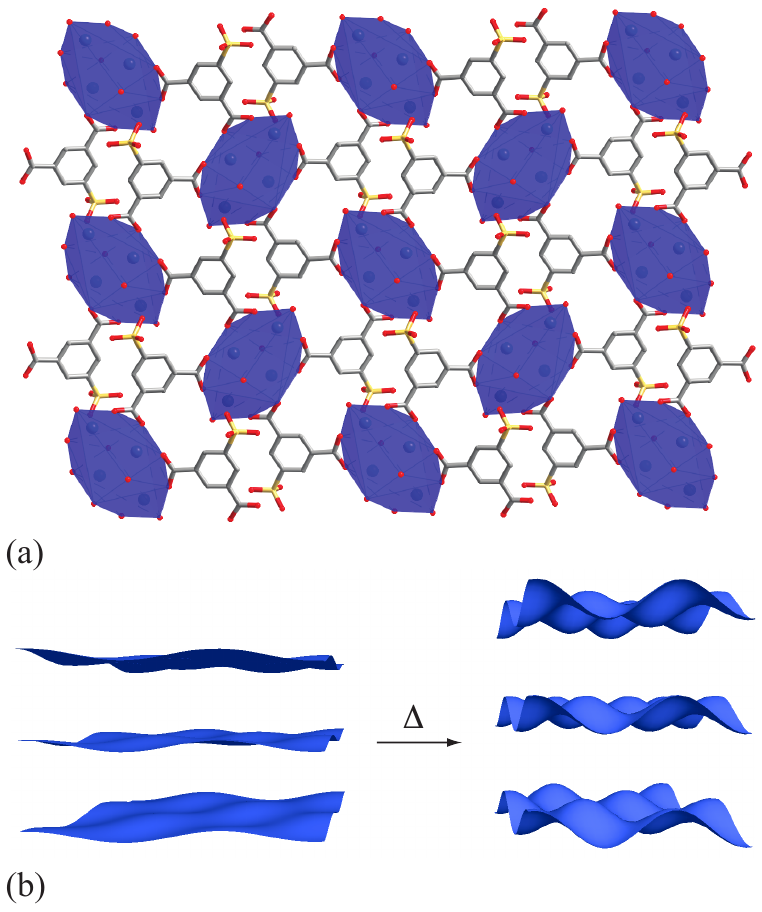}
\end{center}
\end{figure}

The first of our case studies concerns the flexible metal--organic framework Cu$_2$(OH)(C$_8$H$_3$O$_7$S)(H$_2$O)$\cdot$2H$_2$O (abbreviated to Cu-SIP-3$\cdot$3H$_2$O). This is a material of particular currency for its potential applications as a biomedical  agent for controlled nitrous oxide (NO) release \cite{Xiao_2009,Hinks_2010}. Under ambient conditions its structure can be considered as a stack of two-dimensional, covalently-connected `sheets': each such sheet is assembled from what is technically a distorted square (4,4)-net of tetrameric Cu$_4$ clusters connected via 5-sulfoisophthalate linkers (although a description in terms of hexagonal close packing of the same clusters might be considered equally valid) [Fig.~\ref{fig6}(a)]. Adjacent sheets interact only very weakly---via axial Cu$^{2+}\ldots$O$_3$SR contacts ($r($Cu$\ldots$O$)>2.8$\,\AA)---such that if only strong covalent bonds are considered then the network of the solid as a whole has two-dimensional rather than three-dimensional connectivity.

Variable-temperature (150--365\,K) lattice parameter data for this ambient phase were reported by \citeasnoun{Allan_2010} as part of a detailed study into a reversible, reconstructive phase transition that occurs on desolvation/resolvation. Because Cu-SIP-3$\cdot$3H$_2$O crystallises in the monoclinic space group $P2_1/n$, this is an example of a system where not all of the coefficients of thermal expansion determined from lattice parameters correspond to principal expansivities. Nonetheless there are hints even in the lattice parameter measurements that the system exhibits anomalous thermal expansion behaviour: the $b$ unit cell parameter decreases on increasing temperature, and there is a negligible change in unit cell volume over the temperature range studied.

We have used the lattice parameter data of \citeasnoun{Allan_2010} as input for {\sc pasc}al in order to determine the fundamental thermal expansion properties of this interesting material; our results are summarised in Table~\ref{table1}. Perhaps the most unexpected finding here is that two of the three principal expansivities are in fact negative. So not only does Cu-SIP-3$\cdot$3H$_2$O become one of the few known materials to show an area-NTE effect, but this effect is very large indeed. In fact the area coefficient of thermal expansion $\alpha_A=\alpha_1+\alpha_2=-76$\,MK$^{-1}$ is essentially an order of magnitude more negative than corresponding values in previously-studied area-NTE materials: \emph{e.g.}\ the Ni(CN)$_2$ family ($\alpha_A=-11$ to $-16$\,MK$^{-1}$) \cite{Hibble_2007,Hibble_2011}, graphite ($\alpha_A=-2.4$\,MK$^{-1}$) \cite{Bailey_1970} and low-temperature arsenic ($\alpha_A=-4.0$\,MK$^{-1}$) \cite{Munn_1972}.

Clues as to the microscopic driving force responsible for this unusual behaviour can be found in the orientation of the principal axes relative to the crystal structure. The set of directions in which NTE is observed lies perpendicular to the $\mathbf x_3$ principal axis (\emph{i.e.}, the plane containing the $\mathbf x_1$ and $\mathbf x_2$ axes). Hence the orientation of $\mathbf x_3$ must carry some special significance. We find that $\mathbf x_3$ is approximately parallel to the $[101]$ crystal axis, which is in fact the layer-stacking direction (\emph{i.e.} normal to the layer shown in Fig.~\ref{fig6}(a)). Hence, on heating, the framework Cu-SIP-3 expands in a direction parallel to the stacking axis, but contracts in every perpendicular direction.

The general propensity for layered materials to exhibit area-NTE was in fact predicted sixty years ago \cite{Lifshitz_1952}: the lowest-energy phonon branches correspond to `rippling' of the layers, a displacement pattern that causes the interlayer separation to increase and the effective layer area to decrease [Fig.~\ref{fig6}b]. The actual magnitude of NTE response observed then depends on the elastic stiffness of the sheets (\emph{i.e.}, how easily they might be corrugated). We would suggest that the use of flexible organic linkers is responsible for the large NTE effect we find here by virtue of the low energy cost associated with flexing of the layers. Finally we comment that in the light of an emerging correspondence between negative thermal expansion and negative compressibility in framework materials \cite{Goodwin_2008,Fortes_2011,Cairns_2012}, one might expect Cu-SIP-3 to exhibit the elusive property of negative area compressibility under hydrostatic pressure \cite{Baughman_1998}.

\subsection{A molecular crystal: HMX}
Anisotropy in thermal expansion can be of considerable importance in materials applications, particularly when the avoidance of cracking is important. One such application is the development of high-performance explosive materials, where cracking can increase the propensity for accidental detonation---\emph{i.e.} the sensitivity of the explosive \cite{Bennett_1998,McGrane_2009,Ramaswamy_1996}. Like many molecular crystals, explosives often crystallise in low-symmetry crystal systems. Consequently in order to determine the degree of anisotropy in their mechanical response it is rarely sufficient to consider the temperature-dependence of the unit cell lengths alone. Indeed since the principal coefficients of thermal expansion correspond to the extrema of mechanical response, the use of unit cell length variation to characterise mechanical anisotropy will \emph{always} underestimate its maximum extent.

\begin{figure}\label{fig7}
\begin{center}
\caption{(a) A representation of the molecular geometry of HMX in its $\beta$-phase. (b) The thermal expansivity indicatrix viewed down the axis of NTE (\emph{i.e.} for the $\mathbf x_1,\mathbf x_2$ plane). The negative-valued lobe of the expansivity indicatrix is small and projects from the centre of the expansivity indicatrix towards the viewer. (c) A representation of the crystal structure of $\beta$-HMX viewed in the same orientation as for the expansivity indicatrix in (b). A flattening of the central eight-membered C$_4$N$_4$ rings at higher temperatures causes an expansion in the $\mathbf x_1$ and $\mathbf x_2$ directions but a contraction in the $\mathbf x_3$ direction (\emph{i.e.} along the viewing direction).}
\includegraphics{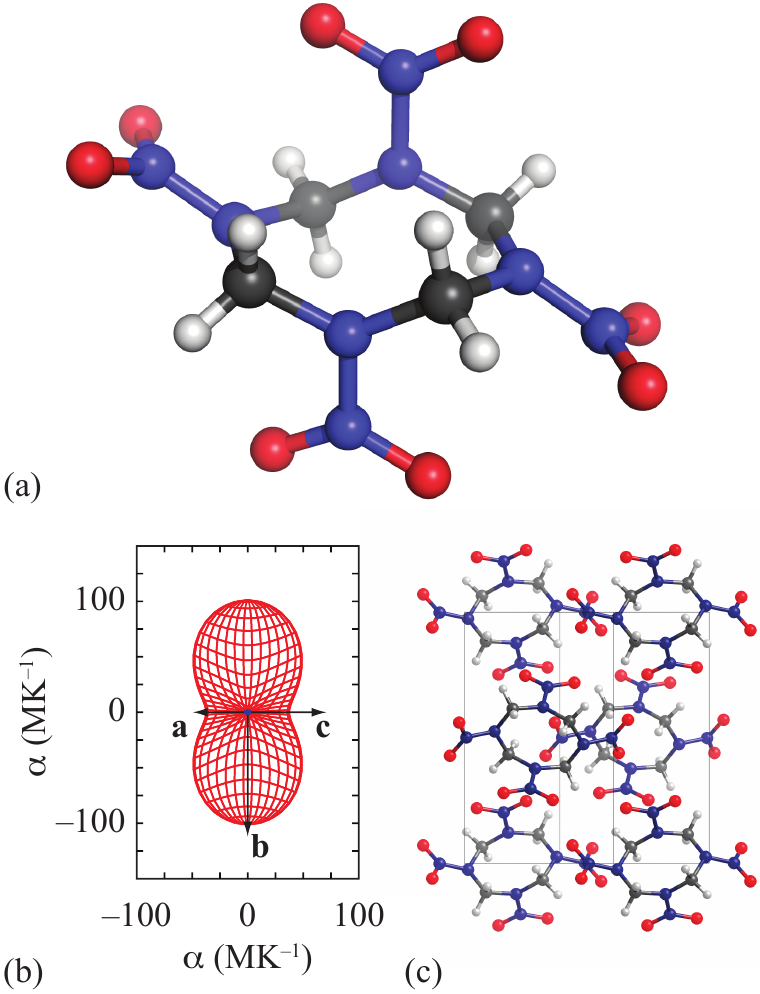}
\end{center}
\end{figure}

One of the most important and widely-used explosives is cyclotetramethylene-tetranitramine (HMX) \cite{Agrawal_2010}. HMX is known to crystallise in any one of four polymorphs ($\alpha, \beta, \gamma, \delta$), of which the $\beta$ phase (space group $P2_1/n$) is the most important because it is the only stable polymorph under ambient conditions [Fig.~\ref{fig7}(a)] \cite{Cady_1963,Choi_1970}. A recent thermal expansion study of $\beta$-HMX, aimed at characterising its mechanical anisotropy, revealed a large difference in the lattice coefficients of thermal expansion: $\alpha_a, \alpha_b, \alpha_c=33, 99$ and $0.43$\,MK$^{-1}$, respectively \cite{Deschamps_2011}. 

Using the same variable-temperature lattice parameter data of \citeasnoun{Deschamps_2011} as input for {\sc pasc}al, we have been able to determine the principal coefficients of thermal expansion; the {\sc pasc}al output is summarised in Table~\ref{table2}. We find a subtle but important difference between the principal and lattice coefficients of thermal expansion: namely that there exists NTE along one of the principal axes ($\mathbf x_3$). This is important because the degree of anisotropy in some directions is now substantially larger than anticipated from the lattice parameters alone; for example, the lattice expansivities underestimate anisotropy in the $(010)$ plane by \emph{ca} 40\%.

At a molecular level, the driving force for the NTE effect we observe appears to be a flattening of the eight-membered ring of each HMX molecule at higher temperatures. There are two different ring orientations within the $\beta$-HMX crystal structure, which are related by the space group symmetry. For each orientation the four ring N atoms are coplanar, so a vector normal to this plane characterises the ring orientation. We find that an average of the two orientation vectors coincides with the NTE principal axis $\mathbf x_3$ to within 1\% [Fig.~\ref{fig7}(b)]. A simple measure of the degree of ring puckering is given by the mean-squared displacement of the four C atoms away from the N-atom plane defined above. Based on the atomic coordinates given by \citeasnoun{Deschamps_2011}, this quantity appears to decrease with temperature---\emph{i.e.} the HMX rings become flatter at higher temperatures---providing a feasible explanation for the reduction in crystallite size in a direction perpendicular to these C$_4$N$_4$ rings.

\subsection{A mineral: malayaite}
Our final case study concerns the mineral malayaite (CaSnOSiO$_4$), which is a calcium tin silicate closely related to titanite (CaTiOSiO$_4$) that---when doped with chromium---finds application as a pink glaze for ceramics \cite{Ingham_1961,Higgins_1976}. From a fundamental science viewpoint, the comparative critical behaviour of the malayaite--titanite family is of general interest in helping understand the interplay between mechanical and electronic properties in framework silicates \cite{Salje_1993,Zhang_1999}. In titanite itself, second-order Jahn Teller distortions of the $d^0$ Ti$^{4+}$ ion lead to local polarisations of the TiO$_6$ octahedra (such as is observed in key ferroelectrics such as BaTiO$_6$ \cite{Shirane_1967}); this effect is reduced on substitution with Sn$^{4+}$. An interest in understanding the effects of pressure and temperature on the interplay between these off-centre distortions and structural transitions has led to a number of variable-temperature and variable-pressure studies on both systems \cite{Kek_1997,Kunz_2000}.

The structure of malayaite consists of chains of vertex-sharing SnO$_6$ octahedra. These chains are connected via SiO$_4$ tetrahedra to form a three-dimensional anionic network structure. Calcium ions occupy a set of voids within this framework such that each Ca$^{2+}$ is `coordinated' by six or seven O atoms (the Ca--O separations being sufficiently large that identification of the coordination sphere is somewhat arbitrary). At a pressure of 4.95(1)\,GPa, malayaite transforms from its ambient monoclinic $A2/a$ phase into a triclinic $P\bar1$ phase \cite{Rath_2003}. The distortion mechanism associated with this transition and also the structural changes observed on further compression within the triclinic phase offer insight into the key dynamical properties of malayaite. So, for example, the high-pressure crystallographic study of Rath {\it et al.} (2003) documents, amongst other things, the effective compressibilities of SiO$_4$ tetrahedra (not very compressible), SnO$_6$ octahedra (more compressible) and CaO$_x$ coordination polyhedra (very compressible); the implication being that vibrational modes resulting in large changes in Ca--O distances play a significant role in the dynamics of ambient-phase malayaite.

Making use of the variable-pressure lattice parameter data reported by Rath {\it et al.} (2003) for the high-pressure triclinic phase of malayaite, we have been able to extend their analysis to the determination of orthogonal strains. Using first a linear fit to these data, {\sc pasc}al obtains the principal compressibilities and principal axes given in Table~\ref{table3}. Of particular note is the NLC behaviour along $\mathbf x_3$---\emph{i.e.}, approximately parallel to the $[\bar441]$ crystallographic axis. We find that the compression mechanism is dominated by concerted rigid-body rotations and translations of the SnO$_6$ and SiO$_4$ polyhedra that cause a contraction within the plane perpendicular to the $[\bar441]$ axis. The polyhedral `rocking' associated with this compression translates to an expansion in a perpendicular direction---the NLC axis. In this process the CaO$_x$ coordination environment undergoes substantive changes in geometry, such that one might suggest the coordination flexibility of Ca$^{2+}$ allows a particularly anisotropic mechanical response to network compression.

\begin{figure}\label{fig8}
\begin{center}
\caption{Compressibility of high-pressure malayaite. (a) Relative changes in the lengths of the principal axes (open circles, solid lines) and $P\bar1$ unit cell axes (crosses and dashed lines) under compression. The variation in unit cell axes is substantially more moderate than the variation in principal axes. (b) The corresponding principal linear compressibilities. The increase in the magnitude of $K_2$ observed at increasing pressure reflects a dynamical instability.}
\includegraphics{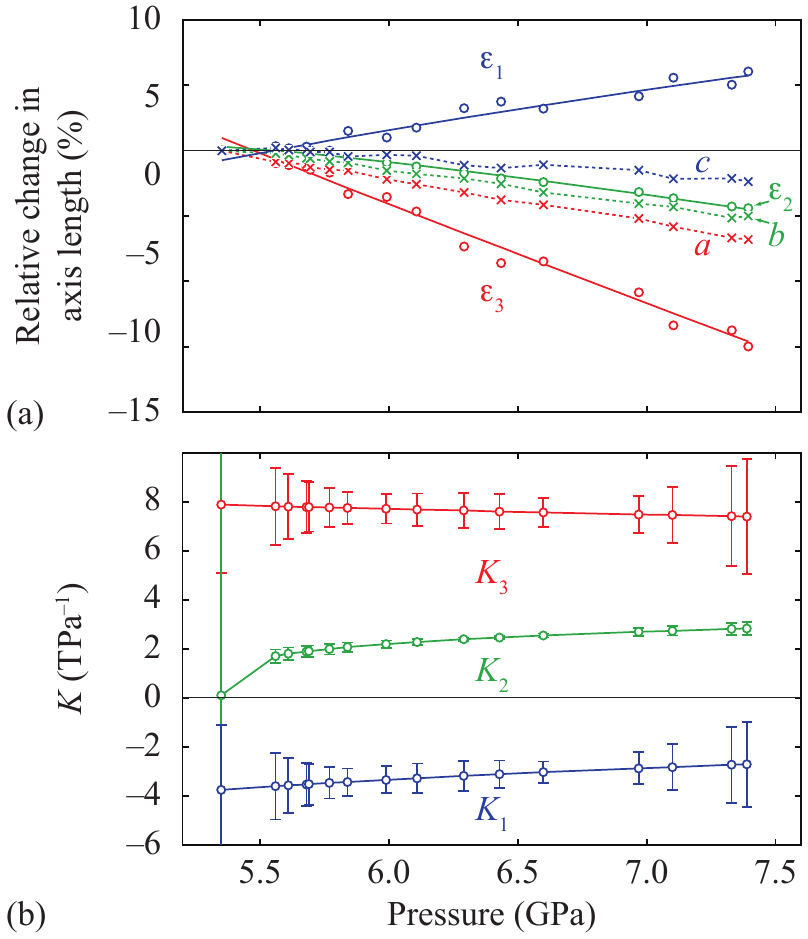}
\end{center}
\end{figure}

For completeness, we show in Fig.~\ref{fig8} the non-linear parameterised fits to the principal strains obtained using {\sc pasc}al and the derived compressibilities. What is apparent from these plots is that the value of $K_2$ increases with pressure---even in this post-transition phase---which could suggest the existence of an incipient dynamic instability at yet higher pressures. This conclusion is reflected also in the thermodynamic quantities obtained from a third-order Birch-Murnaghan fit to the $V(p)$ data: here we find $B_0=170(40)$\,GPa and $B^\prime=-3(5)$. Given the magnitude of the uncertainties involved, we cannot be sure that the value of $B^\prime$ is really negative, but if it were then the implication is that the material is becoming mechanically softer on compression---a feature also associated with dynamic instabilities \cite{Chapman_2007,McConnell_2000}.  To some extent this was discussed by Rath {\it et al.} (2003), but caution led the authors to constrain their fit to a second-order Birch-Murnaghan equation of state. If we perform a similar fit, {\sc pasc}al obtains a value of $B=119(2)$\,GPa that is in close agreement with the value obtained by \citeasnoun{Rath_2003} ($B=118.3(7)$\,GPa). Intriguingly, \citeasnoun{Rath_2002} reports indications of a second high(er)-pressure phase transition in titanite at \emph{ca} 10\,GPa, suggesting that the dynamical anomalies we observe here for malayaite might signal the existence of a similar incipient transition. This putative phase transition could provide an explanation for the observed NLC, as has been observed in other oxide frameworks \cite{Angel_1999,Angel_2004}.

\section{Conclusion}

Our aim has been to develop a simple tool that enables rapid and straightforward calculation of the principal coefficients of thermal expansion and compressibilities from variable-$p/T$ lattice parameter data. On the one hand, this will allow users to screen easily for anomalous mechanical responses---\emph{e.g.}, NTE, NLC and pressure-induced softening. On the other hand, the reporting of principal axes means that any such effects can be related back to the underlying crystal structure. While we are aware that {\sc pasc}al replicates some functionalities of existing codes, it is our hope that the straightforward input format, web accessibility, and ease of output interpretation will mean that a greater scientific audience can make use of orthogonal strain calculations than is currently the case.

The three case studies discussed above illustrate some of the ways in which scientific value might be added through routine lattice parameter analysis with {\sc pasc}al. In the case of the metal--organic framework Cu-SIP-3, we have shown it exhibits the strongest area-NTE effect yet reported, and have been able to relate this effect back to low-energy acoustic modes of its layer-like structure. Analysis of thermal expansion data for the explosive material HMX reveals a much stronger mechanical anisotropy than had otherwise been determined: a result that may have implications for strain propagation and cracking of polycrystalline HMX preparations. Finally, we have shown that high-pressure malayaite exhibits two thermodynamic anomalies: it expands in one direction on compression, and it also becomes mechanically softer on increasing pressure.

These three systems are unlikely to be isolated examples. Indeed, we anticipate that with a more widespread application of principal axis strain calculations, phenomena such as NTE and NLC will likely be observed with increasing frequency. Irrespective of the prevalence of such effects, the mechanisms responsible will almost always provide important insight into the dominant dynamical properties of materials. So it is our particular hope that the ready availability of programs such as {\sc pasc}al might help develop our understanding of dynamics in classes of solid materials for which strain calculations have not yet played a central role.

\ack{Acknowledgements}

The authors gratefully acknowledge financial support from the E.P.S.R.C. (Grant No. EP/G004528/2) and the E.R.C. (Grant No. 279705), and wish to thank D. A. Keen and M. G. Tucker (ISIS) and W. Li (Cambridge) for their assistance in testing our code.

\cleardoublepage

\begin{table}\label{table1}
\caption{Principal coefficients of thermal expansion and corresponding principal axes determined for Cu-SIP-3$\cdot$3H$_2$O.}
\begin{tabular}{cllllc}\hline\hline
&&\multicolumn{3}{c}{Component of $\mathbf x_i$ along the}&\\
Principal&$\alpha_i$ &\multicolumn{3}{c}{crystallographic axes}&\\
axis, $i$&(MK$^{-1}$)&$\mathbf a$&$\mathbf b$&$\mathbf c$&Approximate axis\\[5pt]\hline
1&$-55(7)$&$0$&$-1$&$0$&$[010]$\\
2&$-21(7)$&$-0.7932$&$0$&$0.6090$&$[\bar101]$\\
3&$+60(9)$&$0.8260$&$0$&$0.5637$&$[101]$\\\hline\hline
\end{tabular}
\end{table}

\begin{table}\label{table2}
\caption{Principal coefficients of thermal expansion and corresponding principal axes determined for $\beta$-HMX.}
\begin{tabular}{cllllc}\hline\hline
&&\multicolumn{3}{c}{Component of $\mathbf x_i$ along the}&\\
Principal&$\alpha_i$ &\multicolumn{3}{c}{crystallographic axes}&\\
axis, $i$&(MK$^{-1}$)&$\mathbf a$&$\mathbf b$&$\mathbf c$&Approximate axis\\[5pt]\hline
1&$+101(3)$&$0$&$-1$&$0$&$[010]$\\
2&$+37(3)$&$-0.5570$&$0$&$0.8305$&$[\bar203]$\\
3&$-15(3)$&$-0.8419$&$0$&$-0.5397$&$[302]$\\\hline\hline
\end{tabular}
\end{table}

\begin{table}\label{table3}
\caption{Principal compressibilities and corresponding principal axes determined for the high-pressure ($P\bar1$) phase of malayaite.}
\begin{tabular}{cllllc}\hline\hline
&&\multicolumn{3}{c}{Component of $\mathbf x_i$ along the}&\\
Principal&$K_i$ &\multicolumn{3}{c}{crystallographic axes}&\\
axis, $i$&(TPa$^{-1}$)&$\mathbf a$&$\mathbf b$&$\mathbf c$&Approximate axis\\[5pt]\hline
1&$+7.6(4)$&$-0.6482$&$-0.5429$&$0.5339$&$[\bar1\bar11]$\\
2&$+2.28(10)$&$-0.5846$&$-0.0820$&$-0.8072$&$[203]$\\
3&$-3.1(3)$&$-0.6270$&$0.7591$&$0.1747$&$[\bar441]$\\\hline\hline
\end{tabular}
\end{table}

\cleardoublepage

\cleardoublepage

\cleardoublepage

\cleardoublepage

\cleardoublepage

\cleardoublepage

\cleardoublepage

\cleardoublepage

\cleardoublepage

\end{document}